\documentclass[onecolumn,secnumarabic,amssymb, amsmath, nofootinbib,tightenlines,
nobibnotes, aps, prl,epsfig]{revtex4}
\usepackage{graphicx}
\usepackage{dcolumn}
\usepackage{bm}

\usepackage{amssymb}
\usepackage{epsfig}
\usepackage{color}

\newcommand{\ba}{\begin{eqnarray}}
\newcommand{\ea}{\end{eqnarray}}
\newcommand{\be}{\begin{equation}}
\newcommand{\ee}{\end{equation}}
\newcommand{\bdisplay}{\begin{displaymath}}
\newcommand{\edisplay}{\end{displaymath}}

\begin{document}
\preprint{APS/123-QED}
\title{The ratio of reduced cross-sections in $eA$ processes\\ at Electron-Ion Colliders at $x_{\mathrm{min}}=Q^2/s$}

\author{G.R.Boroun}%
 \email{boroun@razi.ac.ir }
 \affiliation{Department of Physics, Razi University, Kermanshah
67149, Iran}%
\author{B.Rezaei }
\altaffiliation{brezaei@razi.ac.ir}
\affiliation{Department of Physics, Razi University, Kermanshah
67149, Iran}
\date{\today}
\begin{abstract}
We study the predictions of saturation effects in electron-ion
colliders at high inelasticity, using a generalization for nuclear
targets of the ASW and GBW models where the saturation scale,
$Q_{\mathrm{sat}}$, drives the energy dependence and the
corresponding nuclear effects. We expect to observe an enhancement
of the ratio of nuclear reduced cross sections in the saturation
region in future electron-ion colliders. The ratio
$R^{A}_{\sigma_{r}}$ is discussed in the kinematic range of the
electron-Ion collider with center-of-mass energy
$\sqrt{s}=89~\mathrm{GeV}$  at  high inelasticity $y{=}1$. The
importance of the nuclear longitudinal structure function
$F^{A}_{L}$ in the ratio $R^{A}_{\sigma_{r}}$ for the heavy
nucleus lead and light nucleus deuteron at
$x_{\mathrm{min}}=Q^2/s$ is discussed. This enhancement in the
range $Q^2\sim(1-4~\mathrm{GeV}^2)$ in the ratio
$R^{A}_{\sigma_{r}}$ is not observed in the ratio $R^{A}_{F_{2}}$
which is comparable with the nuclear ratio of the nuclear parton
distribution functions. We demonstrate that the study of nuclear
charm structure function allows us to estimate the magnitude of
shadowing effects in high inelasticity in the nuclear gluon
distribution.\\

\end{abstract}
 \pacs{***}
\maketitle
\subsection{I. Introduction}

Distribution of quarks and gluons inside nuclei provides
fundamental insight into the substructure of  hadrons bound in nuclei. The measurement of this distribution is a key goal of the Electron-Ion
Collider (EIC) scientific program \cite{EIC}.  In this program, the
envisioned energy for electron- heavy ion collisions involves colliding electrons with ions at energies of $18~\mathrm{GeV}$ and $110~\mathrm{GeV}$, resulting in a center-of-mass energy of
$\sqrt{s}=89~\mathrm{GeV}$. However, the data from the EIC
program will reach a maximum energy of
$\sqrt{s}{\sim}140~\mathrm{GeV}$, which is lower than the flagship
HERA dataset. In the early 2030s, the EIC at Brookhaven National
Laboratory (BNL) will contribute high -statistical-value data not only on e-p scattering but also on e-A scattering, where A represents the nucleus mass number. The structure of nuclei is of great
interest for  both the EIC and the LHeC \cite{LHeC},  renewing interest in predictions of processes involving nuclei. The EIC will explore the nuclear structure in unprecedented  detail up to the heaviest nuclei, serving as the world$^,$s first eA collider. In
ReF.\cite{Armesto1}, the authors demonstrate the possibility of extracting 
nuclear parton distribution functions (nPDFs) for a single nucleus using EIC pseudodata and fitting techniques similar to those used for
protons. For simulated neutral current (NC) EIC ep
measurements (pseudodata), a grid is produced with five
logarithmically spaced $x$ and $Q^2$ values per decade over the
range of $0.001<y<0.95$. This spacing is well justified by the expected
resolutions in \cite{Armesto1}, where $y$ is the usual inelasticity
variable defined as $y=Q^2/(sx)$.\\
The new data from the EIC or LHeC will illumiate the difference
between the gluon structure of nuclear and proton targets. At low
Bjorken-$x$, the consequence of multiple scattering is known as
shadowing where the nuclear structure function per nucleon is
smaller than that of the proton \cite{Armesto2, Guzey1}. Probing
nuclear structure with the Balitsky-Kovchegov (BK) \cite{BK}
equation with full impact-parameter dependence is considered
in Ref.\cite{Cepila} which extends the study of parton evolution
from proton to nuclear targets. The evolution of the gluon density
in a fastmoving frame is described in Ref.\cite{Guzey2} by
non-linear evolution equations where the nonlinear
Gribov-Levin-Ryskin-Mueller-Qiu (GLR-MQ) \cite{GLRMQ} evolution
equations for nuclear parton distribution function (nPDFs) to
next-to-leading order (NLO) accuracy are studied to quantify the impact of
gluon recombination at small $x$. At a certain scale, gluon recombination, which occurs due to the overlap of gluon fields
from different nucleons, balances splitting, and the gluon density
ceases to grow with increasing interaction energy.\\
Such a regime is referred to as gluon saturation and is characterized
by a saturation scale $Q_{s}^{A}(x)$ which is energy and atomic
number dependent. Perturbative Quantum Chromodynamics (pQCD)
predicts that the small-$x$ gluons in a hadron wavefunction should
form a Color Glass Condensate (CGC). The predictions of CGC
physics for the EIC at high energies are discussed in Ref.\cite{Navar}. The
authors studied how the nucleus at high energies acts as an
amplifier of the physics of high parton densities and estimated the
nuclear DIS structure functions using a generalization for nuclear
targets of the Iancu-Itakura-Munier (IIM) \cite{IIM} model,  which
describes the ep HERA data quite well. The nuclear unintegrated
gluon distribution (nUGD) can be obtained from the nucleon
distribution by using the Glauber-Mueller \cite{GM} approach for
multiple scattering. In this approach,  the dipole scattering matrix in
configuration space, denoted by $r$ \footnote{Where $r$ is the relative
transverse separation between a quark  and an anti-quark in the dipole
model.}, can be determined from the cross section for dipole
scattering off a proton in the following form
\begin{eqnarray}\label{Sdipole_eq}
\sigma_{\mathrm{dip}}^{A}(x,r)=\int d^{2}b~
2\bigg{[}1-\exp\bigg{(}-\frac{1}{2}AT_{A}(b)\sigma_{\mathrm{dip}}^{A}(x,r)\bigg{)}\bigg{]},
\end{eqnarray}
 with $b$ as the impact parameter of the center of the
dipole relative to the center of the nucleus, $T_{A}(b)$ is the
thickness function. This function depends on the impact parameter $b$ and is
normalized to
unity, such that $\int d^{2}b~ T_{A}(b)=1$ \cite{Armesto3}.\\
The differential cross section for lepton-nucleus interactions can be expressed of DIS structure functions through virtual photon exchange, in  the following form
\begin{eqnarray} \label{Dsigma_eq}
\frac{d^{2}\sigma^{lA}}{dxdQ^2}=\frac{4{\pi}\alpha^{2}}{Q^4}\frac{F_{2}^{A}(x,Q^2)}
{x}\left[
1-y-\frac{Q^2}{4E^2}+\frac{y^2+Q^2/E^2}{2(1+R^{A}(x,Q^2))}\right],
\end{eqnarray}
where $\alpha$ is the fine-structure constant and the quantities
$x$ and $Q^2$ are fully determined by the kinematic conditions of
the incident and scattered leptons and the target nuclei.
$F_{2}^{A}$ is the structure function $F_2$ for a nuclear target A
which is found to be different from the mere superposition of A
free nucleon structure functions $F_{2}^{p}$.  $R^A$ is the
ratio of the longitudinal to transverse structure functions, which
gives a small contribution to the cross section.\\
Groups such as Eskola, Kolhinen and Salgado (EKS) \cite{EKS}, de
Florian and Sassot (DS) \cite{DS},  Hirai, S. Kumano and T. H.
Nagai (HKN) \cite{HKN}, and by K. J. Eskola, H. Paukkunen and C. A.
Salgado (EPS) \cite{EPS} are proposed parameterizations of the
nuclear parton distribution functions. The new nCETQ15 extends
CTEQ proton PDFs to include nuclear dependence in \cite{nCTEQ}
using data on nuclei up to $ ^{208}\mathrm{Pb}$ with
uncertainties calculated using the Hessian method.\\
The HIJING \cite{HIJING} parametrization is in good agreement with
the ALICE experiment at LHC energies,  providing a more
stringent constraint on gluon shadowing. The nuclear modification
factors in the HIJING parametrization are as follows:
\begin{eqnarray} \label{RF2_eq}
R^{A}_{F_{2}}(x)=\frac{F_{2}^{A}}{AF_{2}^{p}}= 1+1.19(\ln
A)^{1/6}(x^3-1.2x^2+0.21x)
-s_{q}(b)(A^{1/3}-1)^{0.6}(1-3.5\sqrt{x})\exp(-x^2/0.01),
\end{eqnarray}
and
\begin{eqnarray} \label{RG_eq}
R^{A}_{G}(x)=\frac{xg^{A}}{Axg^{p}}=1+1.19(\ln
A)^{1/6}(x^3-1.2x^2+0.21x)
-s_{g}(b)(A^{1/3}-1)^{0.6}(1-1.5{x}^{0.35})\exp(-x^2/0.004),
\end{eqnarray}
where $s_{i}(b)$ reads
\begin{eqnarray} \label{Sb_eq}
s_{i}(b)=s_{i}\frac{5}{3}(1-b^2/R_{A}^{2}),~~~~~~i=q,g
\end{eqnarray}
with  $s_{q} = 0.1$ and $s_{g}= 0.22-0.23$. The nuclear shadowing
effect is visible over a wide kinematic range
$10^{-5}{\leq}x{\leq}0.1$ and
$0.05{\leq}Q^2{\leq}100~\mathrm{GeV}^2$ from the experimental data
\cite{Arneodo}, which is associated with the modification of the
target parton distributions as $xq^{A}(x,Q^2)<Axq^{p}(x,Q^2)$. The
possibility of constraining the nuclear effects in the parton
structure of nuclei using the inclusive observables, which would
be measured in the EIC, is considered in Ref.\cite{Navarra}.
Nuclear shadowing is controlled by the interplay between the life
time of photon fluctuations (or coherent time where shadowing is
possible only if the coherence time exceeds the mean internucleon
spacing in nuclei) and shadowing saturates if the coherent time
substantially exceeds the nuclear radius. The study of shadowing
for transverse and longitudinal photons based on the results for
DIS off nuclei from the HERMES experiment is done in
Ref.\cite{Kop1}. In Ref.\cite{Kop2}, the authors introduced a
new scaling variable in terms of which nuclear shadowing in DIS is
universal. Prospects for constraining the nuclear distribution
functions by small $x$ DIS by including a sample of pseudodata at
the LHeC collider are studied in Ref.\cite{Eskola}.\\
The purpose of this work is to use well -defined models for
the behavior of gluon density in the DIS structure functions at
small $x$.  In particular, we consider the available models for
the lepton-nucleus interaction and  the dipole cross section which
incorporate the evolved gluon distribution functions. The purpose
of this paper is to evaluate the nuclear reduced cross section in
the kinematic regions corresponding to the EIC for $eA$ collision
at high inelasticity where $x_{\mathrm{min}}=Q^2/s$ . We
produce results for the
$\frac{\sigma_{r}^{A}}{A\sigma_{r}^{p}}(s,Q^2)$ using the
$\gamma^{*}A$ interaction models.\\
The structure of the manuscript is as follows: In Section II we
review the method to compute the nuclear reduced cross section as
the virtual photon-nucleus cross section is smaller than A times
the photon-nucleon cross section. Then, we present our
results and conclusions in Section III.\\

\subsection{II. Method}

The importance of experimental data in the low $Q^2$ region will be evident
in the EIC and LHeC through measurements of the longitudinal structure
function in  protons and nuclei. The determination of the
longitudinal structure function, based on an extrapolation of the
HERA deep inelastic scattering reduced neutral current cross
section data at fixed $\sqrt{s}$ and $Q^2$ to the minimum value of
$x$ given by $Q^2/s$ is considered in Refs.\cite{Taylor, Boroun1}.
The authors have shown that they expect $F_L$ to be small because
its dominant gluon component is strongly suppressed when
 the polarization of the exchanged photon is transverse in that
kinematic region. Now, we apply the method to the nuclear DIS
structure functions as the nuclear reduced cross section
$\sigma_{r}^{A}(s,Q^2)$ at  high inelasticity can be standardly
defined via the structure functions $F_{2}^{A}(s,Q^2)$ and
$F_{L}^{A}(s,Q^2)$ on the nucleus A by the following form
\begin{eqnarray} \label{SigmaA_eq}
\mathrm{lim}_{y{\rightarrow}1}
[\sigma_{r}^{A}(s,Q^2)]=F_{2}^{A}(s,Q^2)-F_{L}^{A}(s,Q^2).
\end{eqnarray}
The shadowing effect of the reduced cross section of nuclei at
this limit is given by
\begin{eqnarray} \label{RsigmaA_eq}
\mathrm{lim}_{y{\rightarrow}1}
\frac{\sigma_{r}^{A}(s,Q^2)}{A\sigma_{r}^{p}(s,Q^2)}=\frac{F_{2}^{A}(s,Q^2)-F_{L}^{A}(s,Q^2)}{A[F_{2}^{p}(s,Q^2)-F_{L}^{p}(s,Q^2)]}
=R^{A}_{F_{2}}[1-F_{L2}]^{-1}+R^{A}_{F_{L}}[1-F_{2L}]^{-1},
\end{eqnarray}
where $R^{A}_{F_{2}}=\frac{F_{2}^{A}}{AF_{2}^{p}}$,
$R^{A}_{F_{L}}=\frac{F_{L}^{A}}{AF_{L}^{p}}$,
$F_{L2}=\frac{F_{L}^{p}}{F_{2}^{p}}$, and $F_{2L}=[F_{L2}]^{-1}$.
We expect that $F_{L2}{\rightarrow}0$ for $Q^2{\rightarrow}0$ as
predicted in \cite{Taylor, Boroun1} based on the HERA data
\cite{HERA} that is required by electromagnetic gauge
invariance. Therefore, we have:
\begin{eqnarray} \label{LowQ2_eq}
\mathrm{lim}_{Q^2{\rightarrow}0}
[R^{A}_{\sigma_{r}}]{\simeq}R^{A}_{F_{2}},
\end{eqnarray}
where $R^{A}_{\sigma_{r}}=\frac{\sigma_{r}^{A}}{A\sigma_{r}^{p}}$.
In order to calculate proton and nuclear reduced cross sections,
the DIS structure functions can be written as
\begin{eqnarray} \label{F2A_eq}
F_{2}^{A}(s,Q^2)= \frac{Q^2}{4{\pi}^2\alpha}\left(
\sigma_{T}^{\gamma^{*}A}+ \sigma_{L}^{\gamma^{*}A}\right)(s,Q^2),
\end{eqnarray}
and
\begin{eqnarray} \label{FLA_eq}
F_{L}^{A}(s,Q^2)= \frac{Q^2}{4{\pi}^2\alpha}
\sigma_{L}^{\gamma^{*}A}(s,Q^2),
\end{eqnarray}
where the subscripts T and L refer to the photon polarization
states in the dipole model. The nuclear dependence of the
$\gamma^{*}A$ cross section is absorbed in the $A$-dependence of
the saturation scale via geometrical scaling where the cross sections
(i.e., $\gamma^{*}p$ and $\gamma^{*}A$) are rather a function of a
single variable $\tau_{A}=Q^2/Q^{2}_{\mathrm{sat},A}$
\cite{Armesto4}. The geometrical scaling phenomenon defines the
nonlinear perturbative QCD (pQCD) approaches for high energy deep
inelastic electron-proton or electron-nucleus scattering. For
electron-ion collisions, $Q^{2}_{\mathrm{sat},A}$ is enlarged due
to the nuclear enhancement as
$Q^{2}_{\mathrm{sat},A}=A^{1/3}Q^{2}_{\mathrm{sat},p}$, where the
nucleon saturation momentum is set to be
$Q^{2}_{\mathrm{sat},p}=(sx_{0}/Q^2)^{\lambda}~\mathrm{GeV}^2$.
The coefficient parameters $x_{0}$ and $\lambda$ are set based on
the results of fits 0($n_{f}=3$) and 1($n_{f}=4$) with respect to
the HERA data in Ref.\cite{Golec1} based on the quark mass because
the photon wave function depends on the mass of the quarks in the
$q\overline{q}$ dipole which modifies the Bjorken variable $x$ in
the dipole cross section by the following form
\begin{eqnarray} \label{Bjorken_eq}
x{\rightarrow}x\left(1+\frac{4m_{f}^{2}}{Q^2}\right)|_{x_{\mathrm{min}}}=\frac{Q^2}{s}+\frac{4m_{f}^{2}}{s}.
\end{eqnarray}
The $\gamma^{*}A$ structure function is obtained from the
corresponding cross section for $\gamma^{*}p$ process, according
to the Armesto-Salgado-Wiedemann (ASW) \cite{Armesto4} in the
following form
\begin{eqnarray} \label{F2A_eq}
F_{2}^{A}(s,Q^2)= \frac{\pi{R_{A}^{2}}}{\pi{R_{p}^{2}}}
\frac{Q^2}{4{\pi}^2\alpha}\sigma_{tot}^{\gamma^{*}p} \left(
\tau_{p}\left[  \frac{\pi{R_{A}^{2}}}{A\pi{R_{p}^{2}}}
\right]^{\Delta} \right)(s,Q^2),
\end{eqnarray}
where $\delta=1/\Delta=0.79{\pm}0.02$. The nuclear radius is given
by $R_{A}=(1.12A^{1/3}-0.86A^{-1/3})~\mathrm{fm}$ and
$\pi{R_{p}^2}=1.55{\pm}0.02~\mathrm{fm}^2$. The form of the single
universal curve for the photoabsorption cross section in
Refs.\cite{Armesto4, Machado1, Armesto5}  where lie as a function of
the scaling variable $\tau_{p}=Q^2/Q^{2}_{\mathrm{sat},p}$ at the
limit $x_{\mathrm{min}}=Q^2/s$ is the following form
\begin{eqnarray} \label{SigmaP_eq}
\sigma_{tot}^{\gamma^{*}p}(\tau_{p})=\overline{\sigma}_{0}
[\gamma_{E}+\Gamma(0,\xi)+\ln(\xi)],
\end{eqnarray}
where $\gamma_{E}$ is the Euler constant, $\Gamma(0,\xi)$ the
incomplete Gamma function and
\begin{eqnarray} \label{Xi_eq}
\xi=\frac{a}{\tau_{p}^{b}}={a}{\left(
{(Q^2)^{1+\lambda}}{(sx_{0})^{-\lambda}} \right)^{-b}},
\end{eqnarray}
with $a=1.868$ and $b=0.746$ were extracted from a fit to
lepton-proton data and the overall normalization was fixed by
$\overline{\sigma}_{0}=40.56~\mathrm{\mu{b}}$. Therefore
$R^{A}_{F_{2}}$ at the limit $x_{\mathrm{min}}$ is found
\begin{eqnarray} \label{RF2A_eq}
R^{A}_{F_{2}}=\frac{\pi{R_{A}^{2}}}{A\pi{R_{p}^{2}}}\frac{[\gamma_{E}+\Gamma(0,\xi_{A})+\ln(\xi_{A})]}{[\gamma_{E}+\Gamma(0,\xi)+\ln(\xi)]},
\end{eqnarray}
where
\begin{eqnarray} \label{XiA_eq}
\xi_{A}=\frac{a}{\tau_{A}^{b}}={a}{\left(
{(Q^2)^{1+\lambda}}{(sx_{0})^{-\lambda}}\left[
\frac{\pi{R_{A}^{2}}}{A\pi{R_{p}^{2}}} \right]^{\Delta}
\right)^{-b}}.
\end{eqnarray}
In pQCD at NLO in the $\overline{\mathrm{MS}}$ scheme, the
longitudinal structure function $F_L$ is in agreement with the
Altarelli-Martinelli \cite{AM} formula and has the neat expression
where
\begin{eqnarray} \label{FLArmesto_eq}
F_{L}(x,Q^2)=\frac{\alpha_{s}(Q^2)}{2\pi}\sum_{k={q,\overline{q}}}e_{k}^{2}\int_{x}^{1}dz\left[\frac{4}{3}f_{k}\left(\frac{x}{z},Q^2\right)+f_{g}\left(\frac{x}{z},Q^2\right)
(1-z)\right],
\end{eqnarray}
where $f_g$ denotes the gluon PDF, $f_{k}$s the corresponding
quark PDFs, and $e_k$ is the charge of a quark of flavor k.
Nuclear effects on $F_{L}$ at small $x$ are discussed in
Ref.\cite{Armesto6} where the results show closely follow those on
the gluon distribution for PB (A=208). In Ref.\cite{Farid} the
longitudinal DIS structure function for a longitudinally polarized
photon in terms of transverse momentum dependent (TMD) quark and
gluon jet fracture functions is discussed. The result for the
single-inclusive jet cross-section in longitudinally polarized DIS
at  leading power (LP) in $P_{\bot}/Q$ (where $P_{\bot}$ is the
transverse momentum) and small $x$ is
\begin{eqnarray} \label{FLFarid1_eq}
F_{L}^{LP}(x,P_{\bot})=\frac{\alpha_{s}(Q^2)C_{F}}{2\pi}\sum_{k={q,\overline{q}}}e_{k}^{2}x\mathcal{F}_{k}(x,P_{\bot})
+\frac{\alpha_{s}(Q^2)}{3\pi}\sum_{k={q}}e_{k}^{2}x\mathcal{F}_{g}(x,P_{\bot}).
\end{eqnarray}
The authors in \cite{Farid} analogous to the
Altarelli-Martinelli (AM) \cite{AM} identity is obtained the
longitudinal structure function at one loop in the collinear
factorization $\alpha_{s}$ power counting by the following form
\begin{eqnarray} \label{FLFarid2_eq}
F_{L}(x,Q^2)=\frac{\alpha_{s}(Q^2)C_{F}}{2\pi}F_{2}(x,Q^2)
+\frac{\alpha_{s}(Q^2)}{3\pi}\sum_{k={q}}e_{k}^{2}xg(x,Q^2),
\end{eqnarray}
where $F_{2}=\sum_{k=q,\overline{q}}e_{k}^{2}xq_{k}$ is in the
leading-order (LO) naive parton model. In Ref.\cite{Machado2}, the
author investigated the longitudinal structure function at fixed
energy in the color dipole picture (CDP) for models considering
parton saturation effects. The model parameterized based on the
IIM dipole cross section \cite{IIM} which the saturation region
($rQ_{\mathrm{sat}}(x)>2$) has the correct functional form and
obtained either by solving the BK equation \cite{BK} or from the
CGC theory \cite{CGC}. With considering the effective anomalous
dimension, where the dipole cross section in the IIM model has the
following form
\begin{eqnarray} \label{IIM1_eq}
\sigma_{\mathrm{dip}}(x,\mathbf{r})=\sigma_{0}n_{0}\left(\frac{\mathbf{r}^{2}Q_{\mathrm{sat}}^{2}}{4}
\right)^{\gamma_{\mathrm{sat}}+\frac{\ln(2/rQ_{\mathrm{sat}})}{k\lambda{r}}}\Theta(\mathbf{r}-R_{\mathrm{sat}})
+\sigma_{0}\left[1-e^{-a\ln^{2}(b\mathbf{r}Q_{\mathrm{sat}})}
\right]\Theta(R_{\mathrm{sat}}-\mathbf{r}),
\end{eqnarray}
with $R_{\mathrm{sat}}=2/Q_{\mathrm{sat}}$, and expanding around
the value $\mathbf{r}^{2}Q_{\mathrm{sat}}/4=1$  with considering
the two first terms in its Taylor series the dipole cross section
has the following form
\begin{eqnarray} \label{IIM2_eq}
\sigma_{\mathrm{dip}}{\approx}\sigma_{0}(1-2/e)+\sigma_{0}(\mathbf{r}^{2}Q_{\mathrm{sat}}^{2}/4e).
\end{eqnarray}
For fixed $x$, the longitudinal structure function at
$x_{\mathrm{min}}$ is obtained by the following form
\begin{eqnarray} \label{IIM3_eq}
F_{L}(x,Q^2){\approx}\frac{Q^2}{4{\pi}^2\alpha}\left[\frac{\alpha{\sum{e_{k}^{2}}}}{\pi}\sigma_{0}\left(1-\frac{2}{e}\right)+\frac{\alpha{\sum{e_{k}^{2}}}}{\pi}
\frac{\sigma_{0}}{e}\left(\frac{Q_{\mathrm{sat}}^{2}}{Q^2}\right)\right],
\end{eqnarray}
which completed by applying an effective anomalous dimension as a
function of $Q^2$ in the GBW, BFKL, and IIM models in
\cite{Machado2}.\\
We return back to the AM equation according to the results in
Ref.\cite{Boroun2} for the longitudinal structure function and
apply the dominance of the gluon distribution at small $x$, we
have
\begin{eqnarray} \label{FLBC_eq}
F_{L}(x,Q^2){\simeq}\frac{1}{3\pi}\left[\frac{6}{5.9}-\frac{2\alpha_{s}}{3\pi}(1-x)
\right]\sum_{k={q}}e_{k}^{2}\alpha_{s}xg(2x,Q^2),
\end{eqnarray}
where the running coupling at the NLO approximation is defined by
the following form
\begin{eqnarray}\label{Alpha_eq}
\alpha_{s}(Q^2)&=&\frac{4\pi}{\beta_{0}{\ln}(Q^2/\Lambda^2)}\bigg{[}
1-\frac{\beta_{1}{\ln}{\ln}(Q^2/\Lambda^2)}{\beta^{2}_{0}{\ln}(Q^2/\Lambda^2)}\bigg{]},
\end{eqnarray}
where $\beta_{0}$ and $\beta_{1}$ are the first two coefficients of
the QCD $\beta$-function:
\begin{eqnarray}
\beta_{0}&=&\frac{1}{3}(11C_{A}-2n_{f})\nonumber\\
\beta_{1}&=&\frac{1}{3}(34C^{2}_{A}-2n_{f}(5C_{A}+3C_{F})).
\end{eqnarray}
The coefficients $C_{F}=\frac{N_{c}^{2}-1}{2N_{c}}$ and
$C_{A}=N_{c}$ are the Casimir operators in the fundamental and
adjoint representations of the $\mathrm{SU(N_{c})}$ color group
with $N_{c}=3$, and $\Lambda$ is the QCD cut-off parameter and has
been extracted from ZEUS data with
$\alpha_{s}(M_{Z}^{2})=0.1166$.\\
The integrated gluon distribution, as defined by the unintegrated
gluon distribution (UGD) equation
\begin{eqnarray} \label{UGD_eq}
\alpha_{s}\mathcal{F}(x,k_{\bot})=\frac{3\sigma_{0}}{4\pi^2}(k_{\bot}^{2}/Q_{\mathrm{sat}}^{2})
\exp(-k_{\bot}/Q_{\mathrm{sat}}^{2}),
\end{eqnarray}
can be expressed in the following form \cite{Golec2}
\begin{eqnarray} \label{Gluon eq}
\alpha_{s}xg(x,Q^2)=\int_{0}^{Q^2}dk^{2}_{t}\mathcal{F}(x,k_{t})=
\frac{3\sigma_{0}}{4\pi^2}Q^{2}_{\mathrm{sat}}\bigg{[}
1-\bigg{(}1+\frac{Q^2}{Q^{2}_{\mathrm{sat}}}\bigg{)}e^{-\frac{Q^2}{Q^{2}_{\mathrm{sat}}}}\bigg{]}.
\end{eqnarray}
Therefore, the longitudinal structure function due to
Eqs.~(\ref{FLBC_eq}) and ~(\ref{Gluon eq}) is found as
\begin{eqnarray} \label{FLLG_eq}
F_{L}(s,Q^2){\simeq}\frac{1}{3\pi}\left[\frac{6}{5.9}-\frac{2\alpha_{s}}{3\pi}(1-\frac{Q^2}{s}-4\frac{m_{f}^2}{s})
\right]\sum_{k={q}}e_{k}^{2}\frac{3\sigma_{0}}{4\pi^2}Q^{2}_{\mathrm{sat}}(s)\bigg{[}
1-\bigg{(}1+\frac{Q^2}{Q^{2}_{\mathrm{sat}}(s)}\bigg{)}e^{-\frac{Q^2}{Q^{2}_{\mathrm{sat}}(s)}}\bigg{]}.
\end{eqnarray}
The longitudinal structure function at $x_{\mathrm{min}}$ with
the application of nuclear effects,  changing
$Q_{\mathrm{sat}}^{2}{\rightarrow}Q_{\mathrm{sat},A}^{2}$ and
replacing the  target area with the coefficient $A^{2/3}$,
is defined by the following form
\begin{eqnarray} \label{FLLGA_eq}
F_{L}^{A}(s,Q^2){\simeq}\frac{1}{3\pi}\left[\frac{6}{5.9}-\frac{2\alpha_{s}}{3\pi}(1-\frac{Q^2}{s}-4\frac{m_{f}^2}{s})
\right]\sum_{k={q}}e_{k}^{2}A\frac{3\sigma_{0}}{4\pi^2}Q^{2}_{\mathrm{sat}}(s)\bigg{[}
1-\bigg{(}1+\frac{Q^2}{A^{1/3}Q^{2}_{\mathrm{sat}}(s)}\bigg{)}e^{-\frac{Q^2}{A^{1/3}Q^{2}_{\mathrm{sat}}(s)}}\bigg{]}.
\end{eqnarray}

Now, let us discuss charm production and its contribution to the
nuclear structure function at  high inelasticity. The charm
component $F_{2}^{c}$ of the structure function at small $x$ in
the H1 and ZEUS collaborations \cite{HZ} has been found to be
approximately $25\%$ fraction of the total \cite{Navarra}. Here,
at the EIC COM energy, at high inelasticity, we aim to find
the ratio  $R^{A}_{F_{2}^{c}}{\equiv}{F_{2}^{cA}}/{(AF_{2}^{c})}$
at $x_{\mathrm{min}}=Q^2/s$ because the charm component is
directly related to the gluon density in a nuclear environment via
the Bethe-Heitler process $\gamma^{*}g{\rightarrow}c\overline{c}$.
The charm component of the nuclear structure function in the
boson-gluon fusion (BGF) process based on the fixed-order
perturbative theory \cite{Moch1} at high inelasticity is given
by \cite{Moch2}
\begin{eqnarray} \label{F2CA_eq}
F_{2}^{cA}(s,Q^2)=\frac{e_{c}^{2}}{\pi}\int_{\frac{Q^2}{s}(1+\frac{4m_{c}^{2}}{Q^2})}^{1}dy{\frac{x}{y^2}}C^{c}_{g,2}\bigg{(}\frac{x}{y},
\frac{m_{c}^{2}}{Q^2}\bigg{)}\alpha_{s}(\mu^2)yg^{A}(y,\mu^2),
\end{eqnarray}
where the coefficient function is given by
\begin{eqnarray} \label{Coeff_eq}
C^{c}_{g,2}\bigg{(}z,
\frac{m_{c}^{2}}{Q^2}\bigg{)}=\frac{1}{2}\{[z^2+(1-z)^2+z(1-3z)\frac{4m_{c}^{2}}{Q^2}-z^2\frac{8m_{c}^{4}}{Q^4}]{\ln}\frac{1+\beta}{1-\beta}+
\beta[-1+8z(1-z)-z(1-z)\frac{4m_{c}^{2}}{Q^2}]\},
\end{eqnarray}
with $\beta^{2}=1-\frac{4zm_{c}^{2}}{Q^2(1-z)}$. The factorization
scale $\mu$ is assumed to be $\mu^2=4m_{c}^{2}$ or
$\mu^2=4m_{c}^{2}+Q^2$ which shows the uncertainties in the QCD
calculations of the ratio of charm structure functions.\\
The results for the ratio of structure functions and the ratio of
the reduced cross sections (i.e., Eq.~(\ref{RsigmaA_eq})) in a
wide range of atomic number A will be
obtained at high inelasticity ($y=1$) with fixed energy according to the EIC COM energy  in the next section.\\

\subsection{III. Results and Conclusion}

Let us now calculate the results  for
the ratio of the DIS structure functions using  the ASW and GBW models, as shown in
Eq.~(\ref{RsigmaA_eq}) for the reduced cross
sections. The kinematic regions at the EIC are proposed with
$\sqrt{s}=89~\mathrm{GeV}$ where the numerical results are
determined at high inelasticity $y{=}1$, where in this region
$x=^{Q^2}/s$. The values of coefficients were taken from
Refs.\cite{Armesto4} and \cite{Golec2} for the active flavor
numbers $n_{f}=3$ and 4. In Fig.1, predictions of
the ratio $F_{L2}=\frac{F_{L}^{p}}{F_{2}^{p}}(s,Q^2)$ for the EIC
COM energy using the ASW and GBW models in a wide range of $Q^2$ are shown.
The predictions in Fig.1 for high inelasticity can be
directly compared to bounds of the CDP, as given in
Refs.\cite{Boroun3, Dieter1, Machado3}. This plot shows some
interesting features of the ratio $F_{L2}(s,Q^2)$ for
determining the DIS structure function \cite{Boroun4} according
to the EIC COM energy in the future. The behavior of the ratio at
low and high $Q^2$ values is symmetrical, as observed for the HERA
data. The results of H1 data \cite{H1} for the ratio of the DIS
structure
functions at $W=230~\mathrm{GeV}$ are shown in Fig.1.\\
 The ratio at
low $Q^2$ values is small, since the polarization of the exchanged
photon is transverse and its dominant gluon component is strongly
suppressed at this kinematic point. The ratio decreases with
increasing $Q^2$ because transverse polarization starts to
contribute more for higher $x$ and $Q^2$. The results at moderate
$Q^2$ increase as the active flavor numbers increase, and the
ratio is comparable with the CDP bounds. Investigating the DIS
longitudinal structure functions of the proton and nuclei is
interesting at upcoming colliders such as EIC, as they are expected
to improve the precision of $F_{L}$ measurements at low and
moderate $Q^2$ at fixed $s$ \cite{Badelek}. Indeed, the ratio is
sharpened by including information on the charm quark with
$n_{f}=4$ as discussed in Ref.\cite{Ewerz} for
$Q^2{\sim}{\mathcal{O}}(10-20~\mathrm{GeV}^2)$. This behavior
accurately affects the ratio of the reduced cross sections in
the next figures according to Eq.~(\ref{RsigmaA_eq}). At these
kinematic points, data on the ratio of $F_{L2}$ significantly
exceeding the bounds would rule out the standard dipole picture,
which provides information on questions like color transparency and
saturation \cite{Ewerz}. The dependencies of the dipole cross
sections in proton and nuclear targets \cite{Farid2} on the impact
parameter shift this sharpened behavior with the increasing of $Q^2$,
where the saturation scale depends on the impact parameter based
on the b-CGC model \cite{bCGC}.\\
\begin{figure}
\includegraphics[width=.55\textwidth]{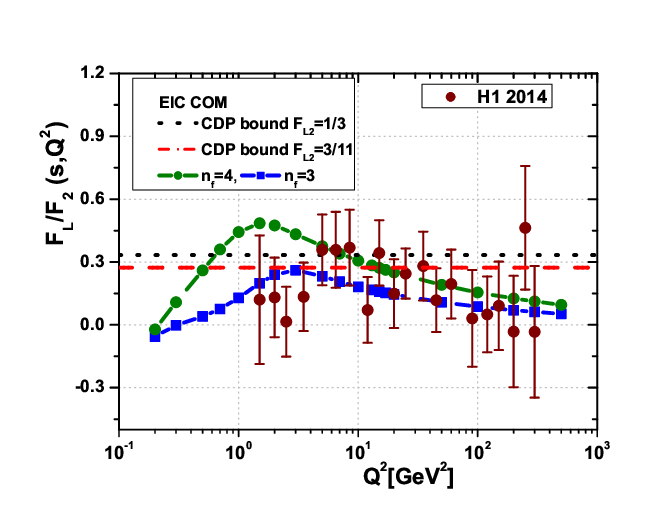}
\caption{The ratio of the DIS structure functions is shown as a
function of $Q^2$ at  high inelasticity $y=1$ for
$x_{\mathrm{min}}=Q^2/s$ according to the EIC COM energy
$\sqrt{s}=89~\mathrm{GeV}$ for $n_{f}=3$ (blue-square curve) and
$n_{f}=4$ (green-circle curve) and compared to the CDP bounds with
$F_{L2}=1/3$ (dot-black line)  and $F_{L2}=3/11$ (dashed-brown
line). The results of H1 data \cite{H1} for the ratio  at
$W=230~\mathrm{GeV}$ are shown  accompanied by the total
uncertainties. }\label{Fig1}
\end{figure}
Nuclear effects on the ratio
$R^{A}_{F_{L}}=\frac{F_{L}^{A}}{AF_{L}^{p}}$  for Pb-208 at the
EIC COM energy at $x_{\mathrm{min}}=Q^2/s$ are shown in Fig.1,
which follows gluon distributions in the GBW model (i.e.,
Eq.~(\ref{FLLGA_eq})). The ratio is independent of the active
flavor number at low and large $Q^2$ values. At large $Q^2$ values
(i.e., $Q^2{\gtrsim}20~\mathrm{GeV}^2$) we observe that
$F_{L}^{A}=AF_{L}^{p}$, and at low values of $Q^2$ (i.e.,
$Q^2{\lesssim}0.8~\mathrm{GeV}^2$) the ratio
$R^{A}_{F_{L}}{\rightarrow}0$ as predicted according to the
polarization of the virtual photon in this region. Thus, a measurement of
the ratio $R^{A}_{F_{L}}$ offers the possibility of quantifying
the nuclear effects on the ratio of gluon distributions at small
$x$ \cite{Armesto6, Bita}.\\
\begin{figure}
\includegraphics[width=.55\textwidth]{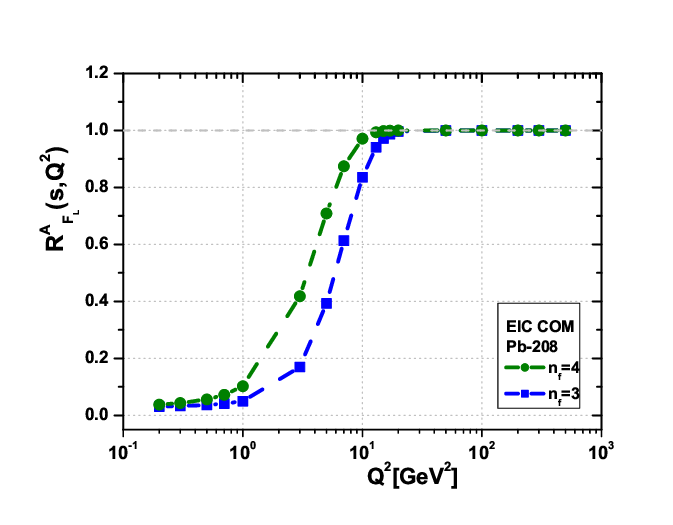}
\caption{Results for $R^{A}_{F_{L}}=\frac{F_{L}^{A}}{AF_{L}^{p}}$
for Pb-208 are shown as a function of $Q^2$ at  high
inelasticity $y=1$ for $x_{\mathrm{min}}=Q^2/s$ according to the
EIC COM energy $\sqrt{s}=89~\mathrm{GeV}$ for $n_{f}=3$
(blue-square curve) and $n_{f}=4$ (green-circle curve).
}\label{Fig2 }
\end{figure}
The importance of the longitudinal structure function for the
proton and nuclear targets will be evident at the LHeC and EIC COM
energies respectively. The longitudinal structure function is
expected to be measured for the first time in the kinematic
regime of small $x$ since the electron-ion collider will be able
to vary the energies of both the electron and ion beams. The
behavior of the ratios $F_{L2}$ and $R^{A}_{F_{L}}$  dominates
the ratio $R^{A}_{\sigma_{r}}$, Eq.~(\ref{RsigmaA_eq}), at low
$Q^2$ values and high inelasticity in the EIC COM energy for A=2
and A=208 in Figs.3 and 4 respectively. In Fig.3, the behavior of
the ratio $R^{A}_{\sigma_{r}}$ is compared with the ratio
$R^{A}_{F_{2}}$ for deuterium with $A=2$ at the EIC COM energy for
$x_{\mathrm{min}}=Q^2/s$. The behavior of the ratio
$R^{A}_{F_{2}}$ (the right panel of Fig.3) as predicted at the
high inelasticity for the EIC COM energy  relies on the nPDFs
\cite{EKS, DS, HKN, EPS, Eskola, Khanpour} at the active flavors
$n_{f}=3$ and $4$. We observe that the effectiveness of the
longitudinal structure function is observable in the ratio
$R^{A}_{\sigma_{r}}$ (the left panel of Fig.3) at low $Q^2$ values
and increases with the increasing number of active flavors, as
observable in the ratio $F_{L2}$ in Fig.1. In particular, the
presence of the ratio $F_{L2}$ in $R^{A}_{\sigma_{r}}$ directly
implies this enhancement in $R^{A}_{\sigma_{r}}$.
\begin{figure}
\includegraphics[width=.65\textwidth]{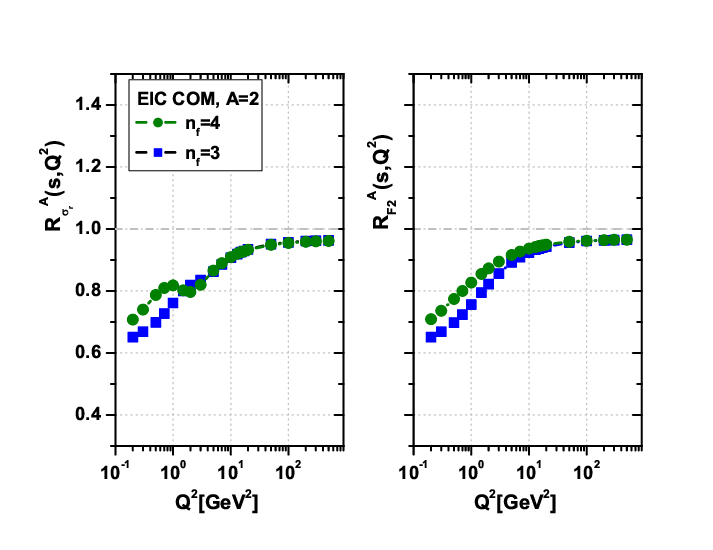}
\caption{Ratios $R^{A}_{\sigma_{r}}$ (the left panel) and
$R^{A}_{F_{2}}$ (the right panel) for deuterium with $A=2$  are
shown as a function of $Q^2$ at high inelasticity $y=1$ for
$x_{\mathrm{min}}=Q^2/s$ according to the EIC COM energy
$\sqrt{s}=89~\mathrm{GeV}$ for $n_{f}=3$ (blue-square curve) and
$n_{f}=4$ (green-circle curve).}\label{Fig3 }
\end{figure}
Conversely, if we assume the nonexistence of the longitudinal
structure function in the ratio of the reduced cross sections in
the nuclear target, no enhancement is present in
$R^{A}_{\sigma_{r}}$ in this kinematical region which is
observable in the ratio $R^{A}_{F_{2}}$ in Fig.3. The enhancements
are observable in the ranges
$1{\lesssim}Q^2{\lesssim}2~\mathrm{GeV}^2$ and
$3{\lesssim}Q^2{\lesssim}4~\mathrm{GeV}^2$ due to the active
flavor numbers $n_{f}=4$ and $n_{f}=3$ in the ratio
$R^{A}_{\sigma_{r}}$ respectively in the EIC COM energy at the
high inelasticity. In Fig.4, these enhancements in the shadowing
region are visible with an increasing  mass number $A$ for
lead with A=208 in the ratio $R^{A}_{\sigma_{r}}$. Therefore, it
suggests that the effects of the longitudinal structure function
in the ratio $R^{A}_{\sigma_{r}}$ can be easily constrained at
$eA$ scattering performed at Brookhaven National Laboratory(BNL)
and Thomas Jefferson National Accelerator Facility (JLab)
\cite{RHIC}. As the small-$x$ region at eRHIC will be probed at
small-$Q^2$ with the characteristic value of
$Q^2=2.5~\mathrm{GeV}^2$, the shadowing effects before the
enhancements will be visible for lead but similar conclusions are
obtained for other values of the atomic number. Consequently, by
measuring $R^{A}_{F_{2}}$ it is possible to constrain the
existence and magnitude of the shadowing effects at
$x_{\mathrm{min}}=Q^2/s$ according to the EIC COM energy
$\sqrt{s}=89~\mathrm{GeV}$.\\
\begin{figure}
\includegraphics[width=.55\textwidth]{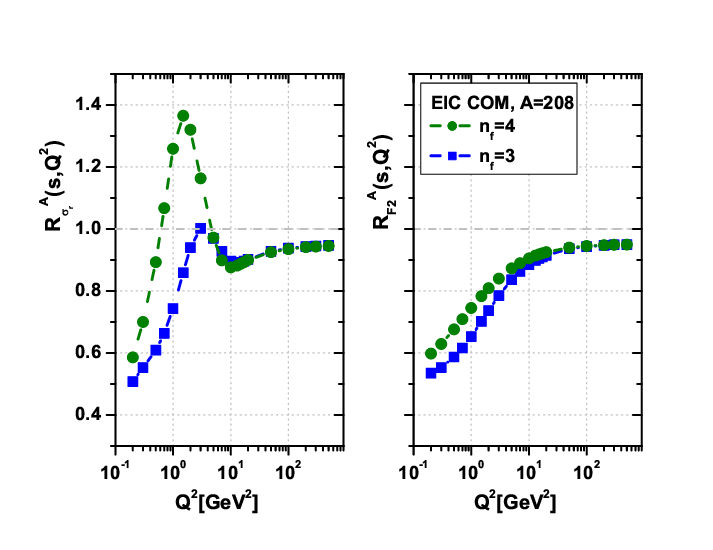}
\caption{The same as Fig.3 for lead with $A=208$. }\label{Fig4 }
\end{figure}
The importance of the gluon density in a nuclear environment is
considered in the ratio $R^{A}_{F_{2}^{c}}$ in Fig.5. In this
figure, we consider the ratio ${F_{2}^{cA}}/{(AF_{2}^{c})}$ at
$x_{\mathrm{min}}=Q^2/s$ as a bound in the EIC COM energy, which is
similar to the parameterization groups. In Fig.5, we consider the
ratio for $A=208$ in a wide range of $Q^2$ values. We observe that
$R^{A}_{F_{2}^{c}}$ suggests an upper bound for the magnitude of
the  shadowing effects at the EIC COM energy at the high
inelasticity. The behavior of the ratio at the renormalization
scale $\mu^2=4m_{c}^{2}$ with the gluon distribution from the GBW
model \cite{Golec2} is similar to the nuclear gluon distributions
from DS and HKN parametrizations \cite{Navarra}. The result at the
scale $\mu^2=Q^2+4m_{c}^{2}$ shows an enhancement in
$R^{A}_{F_{2}^{c}}$  in the interval
$1{\lesssim}Q^2{\lesssim}10~\mathrm{GeV}^2$ which does not
imply the antishadowing effects where nonexistence in the
nuclear gluon distribution. At the scale $\mu^2=Q^2+4m_{c}^{2}$
and $\mu^2=4m_{c}^{2}$  for $Q^2{\gtrsim}10~\mathrm{GeV}^2$ and
$Q^2{\gtrsim}100~\mathrm{GeV}^2$ respectively, we observe that
$F_{2}^{cA}=AF_{2}^{c}$ which indicates the emergence of the
saturation regime of QCD.\\
\begin{figure}
\includegraphics[width=.55\textwidth]{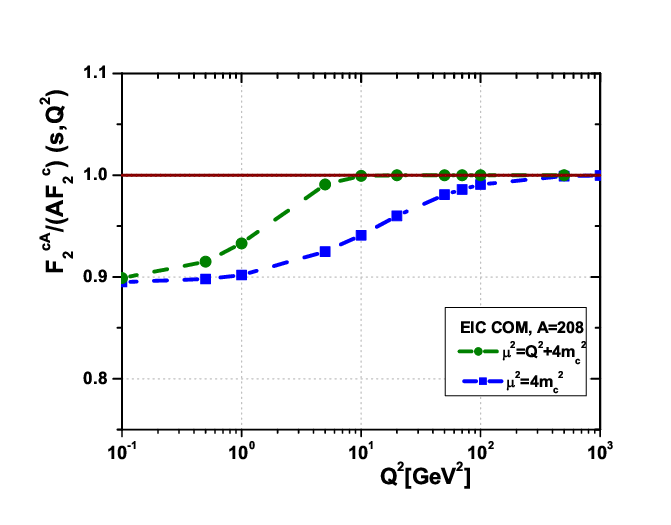}
\caption{Ratio $R^{A}_{F_{2}^{c}}$  for lead with $A=208$  is
shown as a function of $Q^2$ at high inelasticity $y=1$ for
$x_{\mathrm{min}}=Q^2/s$ according to the EIC COM energy
$\sqrt{s}=89~\mathrm{GeV}$ for $\mu^2=4m_{c}^2$ (blue-square
curve) and $\mu^2=Q^2+4m_{c}^2$ (green-circle curve). }\label{Fig5
}
\end{figure}

As a summary, in this paper, we have studied the predictions of
saturation physics for electron-ion colliders at high inelasticity,
where $x_{\mathrm{min}}=Q^2/s$, using a generalization of the GBW and ASW models for the DIS structure functions
on nuclear targets. These models accurately describe the ratio
$F_{L2}=\frac{F_{L}^{p}}{F_{2}^{p}}(s,Q^2)$ observed in ep collisions at HERA.  We have estimated the ratio of nuclear structure
functions $R^{A}_{F_{2}}$, which is comparable to the nPDFs in
the literature, showing shadowing effects at $x_{\mathrm{min}}$
and non-linear corrections at low $Q^2$ values due to the EIC COM
energy. Additionally, we have demonstrated  an enhancement behavior
in the ratio of the nuclear reduced cross section
$R^{A}_{\sigma_{r}}$ on the order of
${\mathcal{O}}(1-4~\mathrm{GeV}^2)$ due to the active flavor
numbers at the EIC COM energy for deuteron
and lead targets. These enhancements are indicative of  the saturation regime, driven by
the behavior of the ratio of longitudinal structure functions in
$R^{A}_{\sigma_{r}}$. These results suggest that studying the
ratios $R^{A}_{\sigma_{r}}$, $R^{A}_{F_{2}}$, $R^{A}_{F_{L}}$, and
$F_{L2}$ in EIC and LHeC colliders is ideal for constraining
nuclear effects on the nuclear gluon distribution.Furthermore, by measuring these observables, we can directly
access the nuclear gluon distribution in future accelerator experiments. We hope that this
paper will encourage a more precise determination of the ratio
$R^{A}_{\sigma_{r}}$ and underscore the importance of  nuclear longitudinal
structure functions at low-$x$ and low-$Q^2$ values. The nuclear
effects on the nuclear gluon distribution are further examined in
the ratio $R^{A}_{F_{2}^{c}}$ at high inelasticity $y=1$ for
$x_{\mathrm{min}}=Q^2/s$ at to the EIC COM energy of
$\sqrt{s}=89~\mathrm{GeV}$ for  renormalization scales
$\mu^2=Q^2+4m_{c}^2$ and $\mu^2=4m_{c}^2$, illustrating the
saturation regime of QCD at moderate and high $Q^2$ values,
respectively. Shadowing effects are evident at low $Q^2$
values, particularly at very low $x$ values at the
EIC COM energy. It is our hope that this paper will inspire a more precise determination of $F_{L}^{A}$, $\sigma_{r}^{A}$, and
$F_{2}^{cA}$ in the coming years at very low $x$ values in future colliders.\\

\subsection{ACKNOWLEDGMENTS}

The authors are thankful to Razi university for financial support
of this project. G.R.Boroun would also like to express gratitude to Professor Nestor Armesto for his helpful comments and invaluable support.\\





\end{document}